\documentclass[twocolumn,floatfix, showpacs, nofootinbib]{revtex4}

\usepackage[final]{epsfig}
\usepackage{amsmath}
\usepackage{parskip}
\usepackage[dvips]{color}

\definecolor{Joerg}{rgb}{0.0, 0.0, 1.0}

\begin{document}

\title{What the NA60 dilepton data can tell}

\author{Thorsten Renk}
\email{trenk@phys.jyu.fi}
\affiliation{Department of Physics, PO Box 35 FIN-40014 University of Jyv\"askyl\"a, Finland}
\affiliation{Helsinki Institut of Physics, PO Box 64 FIN-00014, University of Helsinki, Finland}
\author{J\"org Ruppert}
\email{ruppert@phy.duke.edu}
\affiliation{Department of Physics, Duke University, Box 90305,  
             Durham, NC 27708, USA}


\pacs{25.75.-q}
\preprint{DUKE-TH-????}

\begin{abstract}
The NA60 experiment has studied low-mass muon pair production in In-In collisions at 158 AGeV with unprecedented precision. Thus, there is hope that the in-medium modifications of the vector meson spectral function can be constrained more thoroughly than from previous experiments. Towards this goal, we present a study comparing different models for the in-medium modification of vector mesons. In particular, we investigate what can be learned about collisional broadening
by a medium at finite temperature and baryon density and what model independent
constraints can be obtained from the data. Of special relevance is a comparison
to $p_T$-cuts of the dilepton spectra. Those not only provide independent tests
of the evolution model and $p_T$-dependence of the spectral function but also
unambiguously decide if the observed yield in the  higher mass spectrum $M > 1 {~\rm GeV}$ is mainly due to the hot phase of the evolution or four pion annihilation processes.
\end{abstract}

\maketitle

\section{Introduction}
\label{sec_introduction}

The properties of vector mesons, especially the $\rho$-meson, 
are predicted to change in the presence of a hot and dense nuclear medium. The CERES and HELIOS-3 collaborations demonstrated that central nucleus-nucleus collisions result in  a strong enhancement of the emission of low-mass dileptons as compared to scaled proton-nucleus and proton-proton collisions \cite{Masera:1995ck,Agakishiev:1995xb,Agakishiev:1998vt,Lenkeit:1999xu}.

One explanation of these experimental findings is based on broadening of the $\rho$-meson due to interactions with the surrounding medium. Considerable broadening due to the mesonic and the baryonic component as a function of temperature and density has been found in various hadronic model calculation, 
for reviews see e.g \cite{Rapp:1999ej, Alam:1999sc, Gale:2003iz}. 
We will focus here on three specific calculations, namely the application of thermal field theory in a chiral framework \cite{RSW}, a mean field model for hot and dense
nuclear matter \cite{Renk:2003hu} and a non-perturbative $\Phi$-functional approach 
\cite{RR}.

The mass resolution of the dilepton spectra of the CERES experiment was not high enough to conclusively decide which of these scenarios is realized in nature. Unfortunately, not even the seemingly straightforward question whether a mass shift of the $\rho$-meson due to the Brown-Rho scaling law is realized in nature or not could be answered conclusively \cite{Rapp:1999ej}.

The high precision di-muon data of the NA60 experiment \cite{NA60data}  now provide a possibility to put strong constraints on models of the in-medium effects on vector mesons. Towards this aim, we conduct a comparison of the different model calculations with the data.

Our strategy is as follows: Based on an evolution model successfully used to describe a plethora of observables in Pb-Pb collisions \cite{SPS}, we rely on scaling arguments to make a sophisticated attempt at describing the evolution of matter in In-In collisions. Ideally, one would need to tune this evolution scenario to hadronic observables, however in practice these are not yet sufficiently well determined experimentally. We will address an estimate of the error introduced by our lack of knowledge of the underlying evolution later on. 

Using this evolution model, we then calculate the resulting dilepton spectrum for different assumptions of in-medium vector meson properties. A first test of the evolution model
is the comparison with the integrated $p_T$ data. We also show results for
low and high $p_T$ cuts. To the extent that they are reflected in the data, they will lend  a posteriori further confidence to the scaling laws used to determine the evolution.

We argue that different regions of $p_T$ in the di-muon spectra probe different physics, because the $p_T$ dependence of the observed spectrum is in a characteristic way sensitive to the evolution of the flow profile and temperature. 

Studying different regions of $p_T$ also provides the possibility to decide if 
the observed excess in the mass region ($0.9 <M <1.5 {~\rm GeV}$) is mainly
due the emission from a hadronic gas (as suggested in \cite{vanHees:2006ng}) in later stages at lower temperatures (accompanied by higher flow) or from early stage radiation from a hot medium (accompanied by lower flow). Our calculation of the medium evolution suggests that those contribution are predominantly due to the early times of the medium evolution whereas the authors in \cite{vanHees:2006ng} argue that four pion annihilation processes are required to account for the data. 
Since emission from this
temperature region is greatly effected by flow, considerable differences
in the predictions of the $p_T$-dependence of the spectra in both scenarios are expected.


The paper is organized as follows:
In the next section we discuss how the dilepton spectrum is calculated. We especially explain how we employ scaling arguments to readjust a model that was able to 
describe hadronic transverse mass spectra and HBT correlation measurements and dileptons in $158 {~\rm AGeV}$ Pb-Pb and Pb-Au collisions at SPS to the In-In system.
We then briefly discuss the calculations that were employed to descripe the spectral function in the regime $T<T_c$ and $T>T_c$, respectively. We proceed to 
explain the origin of $p_T$ dependence and the technique employed to account for
flow effects on the vacuum $\rho$ signal.
We present the results of our model calculation and furthermore 
discuss test calculations that illustrate the role of the evolution model. Ther we also present the physical mechanism mainly responsible for the high mass ($M \approx 0.9-1.5 {~\rm MeV}$) excess. We conclude the paper with a general discussion of our results.
 
\section{Calculating the dimuon spectrum}

The emitted spectrum of dileptons from an evolving thermalized system \cite{SchenkeNOTE} can be found from the convolution, schematically: 

\begin{eqnarray}
\label{E-1}
\frac{d^3N}{dM dp_T d\eta} = \text{evolution} \otimes \frac{dN}{d^4 x d^4q} \otimes \text{acceptance}\,\,,
\end{eqnarray}

where the rate calculated from the averaged virtual photon spectral function $R(q, T, \rho_B)$ as

\begin{eqnarray}
\label{E-Rate}
\frac{dN}{d^4 x d^4q}  = \frac{\alpha^2}{12\pi^4} \frac{R(q, T, \rho_B)}{e^{\beta p_\mu u^\mu}- 1}.
\end{eqnarray}

Here, $q$ represent the four-momentum of the emitted muon pair,
$T$ is the temperature of the emitting volume element and $\rho_B$ its baryon density.
The fireball evolution encodes information on the evolution of radiating volume, temperature $T$ and baryon chemical potential $\mu_B$, local transverse flow velocity $v_T$,  longitudinal rapidity $\eta$ and in the late evolution stage chemical non-equilibrium properties such as pion chemical potential $\mu_\pi$ and Kaon chemical potential $\mu_K$.

The acceptance is highly non-trivial in the the case of NA60. Therefore 
for all instances
where we 
compare our calculations with the data, 
a full Monte Carlo simulation of the detector has been carried out by the NA60 collaboration with our calculations as input.

Although a Lorentz-invariant quantity like the invariant mass spectrum $dN/dM$ is in principle independent from the flow profile of the medium, the resulting invariant mass spectrum after acceptance folding is not: The acceptance has $p_T$-dependence and thus the hardening of the momentum spectra due to flow alters the shape of the invariant mass distribution. 
Therefore it is mandatory that both the medium evolution and the acceptance are implemented in a more than just schematic way in order to allow for a meaningful comparison of theoretical spectral functions with the data.

Every measurement of the dilepton spectrum always probes a combination of evolution and spectral function as apparent from Eq.~(\ref{E-1}). It is therefore necessary to determine the evolution as reliably as possible in order to conclusively address the question of the behavior of the in-medium spectral function $R(q,T,\rho_B)$. We now discuss each of the ingredients in turn.

\subsection{The evolution model}

A detailed description of the model used to describe the fireball evolution is found in \cite{SPS}. 
The main assumption for the model is that an equilibrated system is formed
a short time $\tau_0$ after the onset of the collision. Furthermore, we assume that this
thermal fireball subsequently expands isentropically until the mean free path of particles exceeds
(at a timescale $\tau_f$) the dimensions of the system and particles 
move without significant interaction to the detector.

For the entropy density at a
given proper time we make the ansatz 
\begin{equation}
s(\tau, \eta_s, r) = N R(r,\tau) \cdot H(\eta_s, \tau)
\end{equation}
with $\tau $ the proper time as measured in a frame co-moving
with a given volume element  and $R(r, \tau), H(\eta_s, \tau)$ two functions describing the shape of the distribution
and $N$ a normalization factor.
We use Woods-Saxon distributions
\begin{equation}
\begin{split}
&R(r, \tau) = 1/\left(1 + \exp\left[\frac{r - R_c(\tau)}{d_{\text{ws}}}\right]\right)
\\ & 
H(\eta_s, \tau) = 1/\left(1 + \exp\left[\frac{\eta_s - H_c(\tau)}{\eta_{\text{ws}}}\right]\right).
\end{split}
\end{equation}
to describe the shapes for a given $\tau$. Thus, the ingredients of the model are the 
skin thickness parameters $d_{\text{ws}}$ and $\eta_{\text{ws}}$
and the para\-me\-tri\-zations of the expansion of the spatial extensions $R_c(\tau), H_c(\tau)$ 
as a function of proper time.  For a radially non-relativistic 
expansion and constant acceleration we find
$R_c(\tau) = R_0 + \frac{a_\perp}{2} \tau^2$. $H_c(\tau)$ is obtained
by integrating forward in $\tau$ a trajectory originating from the collision center which is characterized
by a rapidity  $\eta_c(\tau) = \eta_0 + a_\eta \tau$
with $\eta_c = \text{atanh } v_z^c$ where $v_z^c$ is  the longitudinal velocity of that 
trajectory. Since the relation between proper time as measured in the co-moving frame 
and lab time is determined by the rapidity at a given time, the resulting integral is
in general non-trivial and solved numerically (see \cite{SPS} for details).
$R_0$
is determined in overlap calculations using Glauber theory. $\eta_0$ is the initial size of the rapidity interval occupied
by the fireball matter.  $a_\eta$ and $a_\perp$ are free parameters, but we choose to use the transverse velocity  $v_\perp^f = a_\perp \tau_f$ and
rapidity at decoupling proper time $\eta^f = \eta_0 + a_\eta \tau_f$ as parameters instead.
Thus, specifying $\eta_0, \eta_f, v_\perp^f$ and $\tau_f$ sets the essential scales of the spacetime
evolution and $d_{\text{ws}}$ and $\eta_{\text{ws}}$ specify the detailed distribution of entropy density.

For transverse flow we assume a linear relation between radius $r$ and
transverse rapidity $\rho = \text{atanh } v_\perp(\tau) =  r/R_c(\tau) \cdot \rho_c(\tau)$
with $\rho_c(\tau) = \text{atanh } a_\perp \tau$.

We allow for the possibility of accelerated longitudinal expansion
which in general implies $\eta \neq \eta_s$ \cite{SPS}.  Here, $\eta = \frac{1}{2} \ln\frac{p_0 + p_z}{p_0 - p_z}$
denotes the longitudinal momentum rapidity of a given volume element. We can parametrize this mismatch between
spacetime and momentum rapidity as
a local $\Delta \eta = \eta -\eta_s$ which is a function of $\tau$ and $\eta_s$.

The model parameters have been adjusted to hadronic transverse mass spectra and HBT correlation measurements in 158 AGeV Pb-Pb and Pb-Au collisions at SPS. The framework was then used to successfully describe photon and dilepton emission and charmonium suppression \cite{SPS}. Since at this point no freeze-out analysis or HBT correlation data are available from NA60 for the In-In system, our strategy is to start with the parameter set derived for Pb-Pb collisions and use geometrical scaling arguments to go to In-In. 

We infer the change in total entropy production from the ratio of charged particle rapidity densities
$dN_{ch}/d\eta$ at 30\% peripheral Pb-Au collisions with $2.1 < \eta <2.55$ measured by CERES \cite{Agakishiev:1995xb,Agakishiev:1998vt,Lenkeit:1999xu} and semi-central In-In collisions at $\eta=3.8$ measured by NA60 \cite{NA60data}, $dN_{ch}^{In-In}/dN_{ch}^{Pb-Pb} = 0.68$.
The number of participant baryons and the effective initial radius (for the sake of simplicity, we map the overlap area in non-central collisions to a circle with the same area) are obtained with nuclear overlap calculations. We assume that the stopping power (which determines the width of the inital distribution of entropy and baryon number in rapidity) scales approximately with the number of binary collision per participant, $N_{bin}/N_{part}$. This determines the scale $\eta_0$. The shape parameters $d_{ws}$ and $\eta_{ws}$ have no great impact on electromagnetic emission into the midrapidity slice, they primarily govern the ratio of surface to volume emission of hadrons. Thus, we leave them unchanged from their value determined in \cite{SPS}.
Under the assumption that the
physics leading to equilibration is primarily a function of incident energy we keep the formation time $\tau = 1$ fm/c as in
Pb-Pb collisions. We stress that the final results exhibit no great sensitivity to the choice of either
$\eta_0$ or $\tau_0$ except in the high $M$, high $p_T$ limit.

The biggest uncertaintly is the choice of the decoupling temperature $T_F$. Due to the fact that the In-In system is smaller than Pb-Pb, we expect a higher decoupling temperature. However, without simultaneous knowledge of HBT correlations and transverse mass spectra an unambiguous answer cannot be obtained. We tentatively choose $T_F = 130$ MeV for the time being. Our lack of detailed knowledge translates primarily into an uncertainty in the normalization of the final results (which depend on the length of emission duration which in turn is increased for a lower decoupling temperature). However we find that our model describes the absolute normalization rather well with this $T_f$, so we do not perform further adjustments.

We fix the flow velocity at $T_C$ by choosing the value of transverse flow obtained at $T = 130$ MeV in our Pb-Pb model in the same centrality class. There is likewise an uncertainty (which is to some degree correlated with the decoupling temperature) since flow determines the slope of hadronic (and to a lesser degree electromagnetic) $p_T$ spectra.

The equation of state in the hadronic phase and the off-equilibrium parameters $\mu_\pi, \mu_K$ are inferred from statistical model calculations as described in \cite{SPS, Hadronization}. The resulting fireball evolution is characterized by a peak temperature of about $250$ MeV, a lifetime of $\sim 7.5$ fm/c and a top transverse flow velocity of $0.55 c$ at decoupling. The scales are quite similar to those obtained in a hydrodynamical evolution model in \cite{Dusling:2006yv} but differ considerably from \cite{vanHees:2006ng}.

\subsection{Hadronic model calculations}

The effective degrees of freedom below the phase transition temperature are color singlet bound states, namely hadrons. The photon couples to the lowest-lying dipole excitations of the vacuum, the hadronic $J^P$=$1^-$ states:
the $\rho$, $\omega$ and $\phi$ mesons and multi-pionic states with the appropriate quantum number.  The dilepton data from CERN-SPS has been the driving force for
theoretical attempts to understand how vector-meson properties are modified in hot and dense strongly interacting matter. 
Especially the $\rho$-meson has been the subject
of many theoretical investigations due to its prominent role for dilepton emission.

Different effective hadronic models and techniques for calculations of the properties of hadronic matter near and below the phase boundaries have been suggested.  For 
reviews of the intense theoretical activities, see e.g. \cite{Rapp:1999ej, Alam:1999sc, Gale:2003iz}. 

Most of those calculations constrained by phenomenological data such as the hadronic and electromagnetic decay widths of the particles predict substantial broadening in matter with comparably small shifts of the in-medium masses. 

We study in the present publication the implications of three different models for the calculation of the observed dilepton spectra and compare their predictions for the changes of the in-medium properties of the vector mesons quantitatively with the NA 60 data below. 

The first model is based on \cite{RSW,SW} where the electromagnetic current-current correlator  has been computed in the so called improved vector meson dominance model combined with the chiral dynamics of pions and kaons.
The irreducible photon self-energy in this model can be directly related to the vector meson self-energy: 
\begin{subequations}
\begin{eqnarray}
{\rm Im} \bar{\Pi} = \sum_V \frac{{\rm Im} \Pi_V(q)}{g_V^2}\left|F_V(q)\right|\,\, ,
\\
F_V(q)=\frac{\left(1-\frac{g}{g_V^0}\right)q^2-m_V^2}{q^2-m_V^2+{\rm i} {\rm Im} \Pi_V(p)} \,\,
, 
\end{eqnarray}
\end{subequations}
here $m_V$ are the in-medium renormalized vector meson masses, $g_V^0$
 is the photon vector meson coupling, and g the coupling of the vector meson
 to the pseudoscalar Goldstone bosons. This equation is valid in the model for vanishing three momentum ${\bf q}$ and the calculation of the electromagnetic current-current correlator is applied in this limit. For the evaluation of finite baryon density effects which are relevant at SPS conditions the results from Klingl, Kaiser, and Weise \cite{Klingl:1997kf} are used. Thermal broadening of the $\rho$, $\phi$, and $\omega$ were calculated by Schneider and Weise as modifications on top of the finite baryon density effects using perturbative methods \cite{SW,RSW}.
 In these calculations it is assumed that the temperature and density dependences of the vector meson self-energy factorize. This amounts to neglecting contributions from pion-nucleon scattering where the pion comes from the heat baths. Formally speaking matrix element contributions such as $\left< \pi N | {\cal T} j_\mu(x) j^{\mu}(x) |\pi N \right>$ are not taken into account. 

A second approach \cite{Renk:2003hu}  is based on a different idea: The method of thermofield dynamics is used to calculate the state with minimum thermodynamic potential at finite temperature and density within the mean field sigma model with a quartic scalar self interaction. The temperature and density dependent baryon and sigma masses are calculated  self-consistently.
The medium modification to the masses  of the $\omega$- and $\rho$-mesons in hot nuclear matter including the quantum correction effects are then calculated in the relativistic random phase approximation. The decay widths for the mesons are calculated from the imaginary part of the self energy using the Cutkosky rule. The model also includes an additional $\rho$ broadening contribution by scattering off by baryons.

We also discuss a third model calculation where we solved truncated Schwinger-Dyson equations derived in a $\Phi$-functional approach incorporating a self-consistent resummation of the $\pi-\rho$ interaction 
in order to determine the thermal broadening of the $\rho$-meson, see \cite{RR} and references therein. This $\Phi$-functional approach self-consistently takes into account  the finite in-medium damping width of the pion in the thermal heat bath which contributes to the broadening of the $\rho$-meson \cite{RR,DS1}. Effects of baryons are not yet included in a self-consistent approximation scheme.

The solution in the $\Phi$-functional approach includes the full momentum dependence of the three-dimensional longitudinal and transverse components of the 
$\rho$-meson's spectral function which is especially important for a comparison with the NA60 data measured at integrated and high $p_t$. 

\subsection{The partonic spectral function}

We describe the thermalized partonic evolution phase by means of a quasiparticle model. Details can be found in \cite{RSW}, here we only summarize the essential findings.

As long as the evolution is in a regime where the thermodynamically active degrees of freedom are quarks and
gluons, the timelike photon couples to the continuum of thermally
excited $q\overline{q}$ states and subsequently converts into a charged lepton pair.

The photon spectral function has been calculated at the one-loop level using
standard thermal field theory methods. The well-known leading-order result
for bare quarks and gluons as degrees of freedom is:
\begin{widetext}
\begin{equation}
\label{E-ImPi}
\begin{split}
R(q^0, {\bf q},T) &= - 3 \sum_{f=u,d,s} \theta(q^2-4m_f^2) e_f^2
\left(1+\frac{2m_f^2}{q^2}\right) \sqrt{1-\frac{4m_f^2}{q^2}}\\
\times&\left(1+2\left[\frac{T}{|{\bf q}|} \frac{1}{\sqrt{1-\frac{4m_f^2}{q^2}}}
\ln \left(\frac{f_D\left(\frac{q_0}{2} -\frac{|{\bf q}|}{2} \sqrt{1-\frac{4m_f^2}{q^2}}\right)}
{f_D\left(\frac{q_0}{2} +\frac{|{\bf q}|}{2} \sqrt{1-\frac{4m_f^2}{q^2}}\right)}\right) -1\right]
\right),
\end{split}
\end{equation}
\end{widetext}
where $q = (q^0, {\bf q})$ is the four-momentum of the virtual photon, $e_f$ the quark electric
charge, $f_D$ the Fermi-Dirac distribution and $m_f$ the quark mass of flavour $f$. 
This result, however, holds only
up to perturbative higher order corrections in $g_s$ that take into account
collective plasma effects.

Let us now assume that a quark quasiparticle couples to a photon
in the same way as a bare quark (a form factor representing the
'cloud' of the quasiparticle could in principle
also be included, but in absence of information about the detailed quasiparticle
structure we ignore this point). For a gas of non-interacting
quasiparticles, the one-loop result for $R$ is
adequate, with input properly adjusted.
In the formalism outlined in \cite{RSW}, all higher order QCD effects manifest
in the thermal quasiparticle masses $m(T)$, the thermal vacuum energy
$B(T)$ and the confinement factor $C(T)$.
Incorporation of the first two features in the calculation is straightforward.
The bare
quark masses in eq.~(\ref{E-ImPi}) simply have to be replaced by the $T$-dependent quasiparticle
masses for each flavour.
The thermal vacuum energy $B(T)$ does not contribute to the dilepton rate.

The mechanism for dilepton production at tree-level
is the annihilation of a $q\overline{q}$ pair into a virtual photon where the
quark lines are multiplied by the distributions $f_D(T)$, giving
the probability of finding a quark or an antiquark in the hot medium.
The incorporation of the confinement factor is hence straightforward: since it reduces the number of thermally active degrees of freedom, it also reduces the dilepton rate by a factor of $C(T)^2$.

Let us note at this point that QGP radiation is expected to be important only in the limit of dileptons at high $M$ and high $p_T$. However, in this limit (which corresponds to the high temperature phase of the evolution) the quasiparticle description is not very different from the free quark description, thus we expect the overall uncertainty resulting from the quasiparticle treatment of the QGP to be small.

\subsection{The origin of $p_T$ dependence}

The transverse momentum dependence of a given dimuon detected at midrapidity arises in this framework for two different reasons. First, the virtual photon created in a $q\overline{q}$ collision or the decay of a vector meson may not be purely timelike but have a finite 3-momentum (with respect to the thermal heat bath) instead. This induces a 3-momentum dependence into the in-medium spectral function, hence we have $R(q_0, {\bf q}, T, \rho_B)$ in general.

However, flow effects alter the momentum dependence. While neither photons nor muons take part in flow themselves, the matter they are emitted from is hadronic and participates in collective motion, hence the local restframe in which virtual photons are emitted is in general not the lab frame. Thus, (neglecting rapidity dependence for a moment) muon pairs emitted from a moving volume element roughly gain an additional momentum ${\bf p}_{\rm flow} \approx M {\bf v_T}$ where $M$ is the invariant mass of the pair and ${\bf v_T}$ the local transverse flow velocity.

For this reason, the experimentally measured $p_T$ does in principle not allow to make conclusions about the 3-momentum of the virtual photon in the local restframe. This is an important point since in several model calculations for the spectral function the simplifying assumption ${\bf q} =0$ has been made. We have checked that this assumption leads to harder $p_T$ spectra and overestimates the yield at high $p_T$ significantly. In this case, only a comparison with the data in the range $p_T < 0.5$ GeV is meaningful. While this doesn't strictly guarantee $p_T \approx 0$ in the local restframe, we may note that for small invariant masses $M < 0.5$ GeV ${\bf p}_{\rm flow} < 250$ MeV which ensures a small momentum ${\bf q}$ for most of the integration over the spacetime integration region. We may also note that in the model masses above 1 GeV are dominated by radiation from hot regions, i.e. early times for which the transverse flow (and hence ${\bf p}_{\rm flow}$) is small. 

We have checked within a scenario with full 3-momentum dependence in the spectral function that the distortion of the lineshape due to neglecting ${\bf q}$ is acceptable (on the order of at most 20\%) as long as the comparison with the data in the region $ p_T < 0.5$ GeV is made, a comparison of these calculations with the full $p_T$ integrated data is however not meaningful at all.

\subsection{The vacuum $\rho$-meson}

\label{S-Vacuum-Rho}

Additionally to the dilepton radiation from the medium (which is dominated by the channel $\pi^+\pi^- \rightarrow \rho^0 \rightarrow \gamma^* \rightarrow \mu^+ \mu^-$ and its in-medium corrections) there is also radiation from electromagnetic decays from $\rho$ mesons outside the medium after the thermalized evolution has ceased, i.e. after the hadronic freeze-out. 

In principle this radiation originates also from $\omega$ and $\phi$ meson vacuum decays, however, due to the small width of these states this contribution is experimentally very well identified and has been subtracted \cite{NA60data}. The vacuum decay of $\rho$ mesons is commonly referred to as 'cocktail $\rho$' and obtained by calculating the produced number of $\rho$ mesons by a statistical model calculation.

In such calculations, the number of $\rho$ mesons is determined under the assumption that the relative abundances of hadron species are given by the thermal expectation value above a temperature $T_h$ (close to the transition temperature $T_C$) but that subsequently only resonance decay (and regeneration) occur. In such a framework, the decay of heavy resonances leads to an overpopulation of pion phase space as compared to the  pion phase space density in chemical equilibrium. 
This can be parametrized by the introduction of a pion chemical potential $\mu_\pi$ (which grows with decreasing temperature) and leads via the tightly coupled $\rho \leftrightarrow \pi\pi$ channel to an effective $\mu_\rho = 2 \mu_\pi$. 

However, while tracking of the resonance decays fixes the number of $\rho$-mesons at thermal decoupling, their momentum spectrum remains unknown. In particular, $\rho$ mesons after chemical decoupling undergo not only resonance decay and regeneration but also participate in the collective expansion of matter. While this effect does not influence the integrated invariant mass spectrum, the acceptance leads to a significant influence of flow on the detected signal. 

Thus, we go beyond the statistical model formulation and compute the full momentum spectrum of $\rho$ mesons after decoupling using the Cooper-Frye formula

\begin{equation}
E \frac{d^3N_{\rho^0}}{d^3p} =\frac{1}{(2\pi)^3} \int d\sigma_\mu p^\mu
\exp\left[\frac{p^\mu u_\mu - \mu_\rho}{T_f}\right]
\end{equation}

with $p^\mu$ the momentum of the emitted $\rho$ and $d\sigma_\mu$ an element of the freeze-out hypersurface (determined by the condition $T=T_F$) just as we would calculate the emission of any other hadron from the medium. We let this ensemble undergo electromagnetic decay with vacuum width and lifetime. In this way, the vacuum $\rho$ signal is influenced by the fireball flow field at the time of decoupling as one would expect.

\section{Results}

\subsection{In-medium modifications from many-body calculations}

\begin{figure*}[!htb]
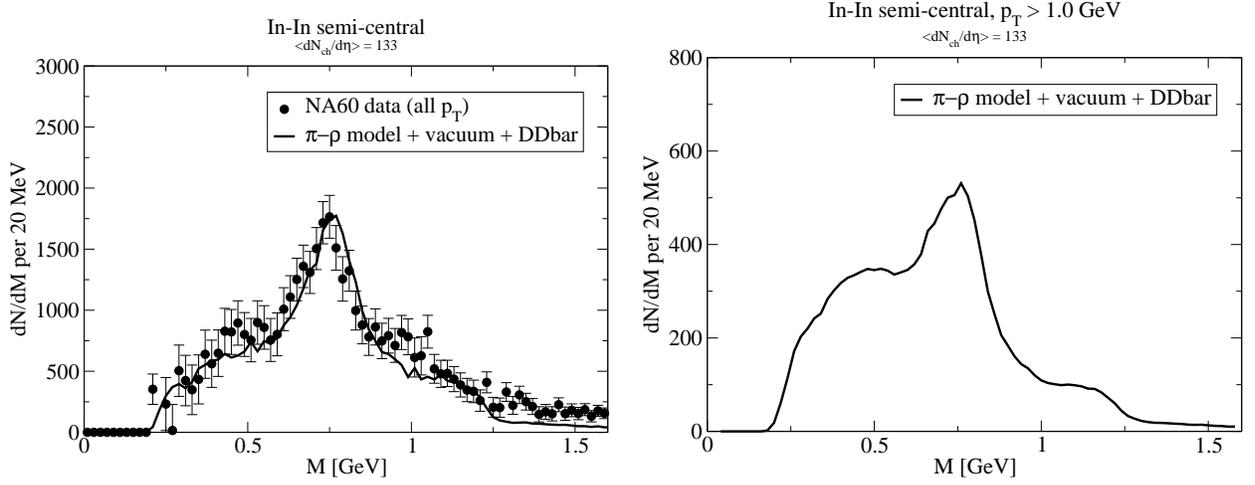

\epsfig{file=dileptons-InIn-RR2-allp_T.eps, width=8cm}
\hspace{2mm}
\epsfig{file=dileptons-InIn-RR2-highp_T.eps, width=8cm}
\caption{\label{F-RR}Left panel: comparison of the NA60 dimuon data 
\cite{NA60data} with calculations within a self-consistent $\Phi$-functional approach for the $\pi-\rho$ interaction \cite{RR}. Right panel: Calculation within a self-consistent $\Phi$-functional approach for the $\pi-\rho$ interaction for a high $p_T$ cut ($p_T>1{~\rm GeV}$).}
\end{figure*}

\begin{figure*}[!htb]
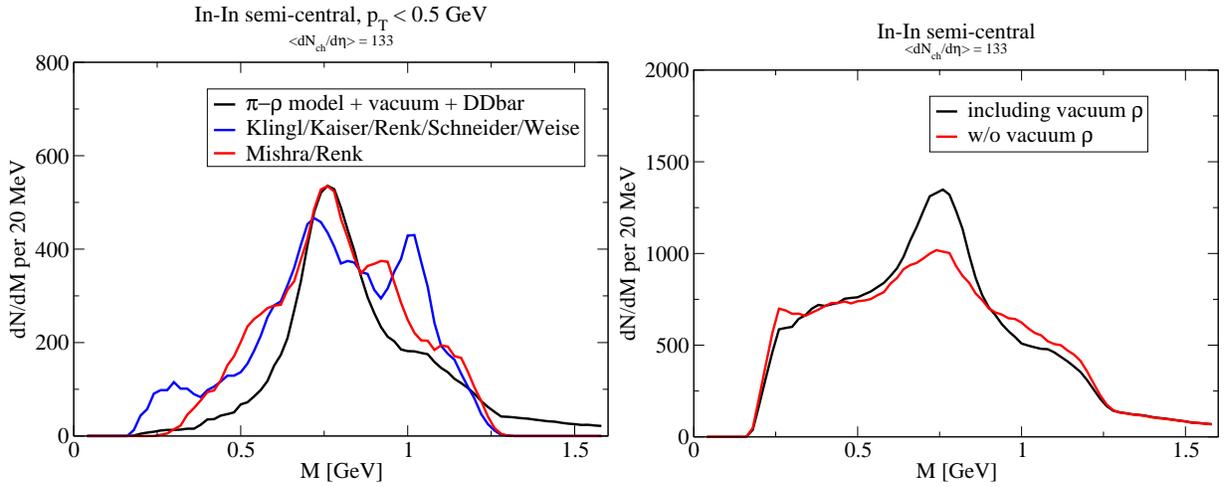

\epsfig{file=dileptons-InIn-RR2-lowp_T.eps, width=8cm}
\epsfig{file=dileptons-InIn-RR2-vacuum_comparison.eps, width=8cm}
\caption{\label{F-LOW}
 Left panel: Calculation of the low $p_T<0.5 {~\rm GeV }$ cut of the dilepton spectrum (detector acceptance of NA60 is taken into account) for the $\Phi$-functional appraoch for the $\pi-\rho$ interaction \cite{RR},  a chiral model thermal field theory ('RSW/KKW') \cite{RSW} for ${\bf q} = 0$, and a thermal nuclear matter calculation ('MR') \cite{Renk:2003hu} for ${\bf q}=0$. Right panel: Comparing the evolution without the vacuum $\rho$ contribution to the full evolution.}
\end{figure*}

We present a comparison of the calculation with the data for $p_T$ integrated data using the spectral function obtained in the $\Phi$-functional approach for the $\pi-\rho$ subsystem \cite{RR} in Fig.~\ref{F-RR}. Furthermore we present calculations
of different muon-pair $p_T$-cuts, i. e. a high $p_T$-cut with $p_T>1 {~\rm GeV}$ in Fig.~\ref{F-RR} and a low $p_T$-cut with $p_T<0.5{~\rm GeV}$ in Fig.~\ref{F-LOW}.

All calculations contain the vacuum $\rho$ calculated using
the Cooper-Frye formula as described in section \ref{S-Vacuum-Rho}. Since 
most of the vacuum $\rho$ contribution is generated during the final breakup of 
the fireball ('volume freeze out') the underlying flow is strong and shuffles strength to higher $p_T$. This is the reason that the vacuum $\rho$ contribution is almost negligible for $p_T<0.5 {~\rm GeV}$ (comparison not shown), but it is important for integrated $p_T$ and $p_T>1 {~\rm GeV}$, see Fig.~\ref{F-LOW}, right panel. 

We note that the relative normalization of vacuum $\rho$ and in-medium $\rho$ appears to  be approximately right. While the strength of the vacuum $\rho$ parametrically scales like the freeze-out volume times the $\rho$ lifetime $V(\tau_f) \tau^\rho_{vac}$, the in-medium radiation scales like $\int V(\tau) d\tau$ and is thus sensitive to the lifetime of the fireball. Thus the good description of the the intermediate mass region $M \approx 0.6 - 0.9~{\rm GeV}$ is due to contributions of the vacuum and in-medium $\rho$ and therefore a consistency test of the fireball model. This  will be even more stringent for the high $p_T$ cut data, because the relative contribution of the vacuum $\rho$ to the in-medium $\rho$ is even more pronounced.

We note that the agreement of the $\Phi$-derivable approach calculation with the data including the vacuum $\rho$ is in very good agreement over the whole range in $M$.  This is to some degree surprising and unexpected, as the model contains substantial amount of the non-perturbative dynamics of the $\pi-\rho$ interaction but is completely insensitive to the presence of other hadrons, especially baryons (which, based on perturbation theory, are expected to play a major role in the broadening of the $\rho$). 

One may take this as an indication that the generic shape of the spectral function and its 3-momentum dependence is right, not a proof that this is the main mechanism capable of producing such a spectral function or that the dynamics of baryons is unimportant. However, this result clearly indicates that it is necessary to investigate non-perturbative mechanisms for broadening in more detail. 

In Fig.~\ref{F-LOW} we also show the results expected from a chiral model calculation \cite{RSW} and a thermal nuclear matter mean field theory calculation \cite{Renk:2003hu} under the assumption ${\bf q}=0$ which restricts the predictive power to the low $p_T$ region. 

The chiral calculation also contains an in-medium $\phi$ and $\omega$ contribution (as these have different mass/width in medium than in vacuum, such  contributions would not be removed by the experimental subtraction of the vacuum $\phi$, $\omega$.)
One interesting finding here is with regard to the $\phi$-meson. Although in the $p_T$ integrated data there appears some room for an contribution in-medium around the $\phi$ mass, no such signal is apparent in the preliminary low $p_T$ data
as presented by S. Damjanovic at Quark Matter 2005 \cite{SanjaQMTalk}.
The KEK-PS E325  collaboration has announced an observation of a considerable in-medium modification of the $\phi$-meson in $12 {~ \rm GeV}$ proton-nucleus collisons that is compatible with the chiral model predictions at low temperature and normal nuclear matter density  \cite{KEK,Klingl:1997kf}. 
Still one has to be careful: in-medium radiation through the $\rho$ channel is primarily 
sensitive to the coupling of $\pi^++\pi^- \leftrightarrow \rho$ whereas radiation from 
the $\phi$ probes $K^+ + K^- \leftrightarrow \phi$ and it is by no means self-evident that both need to be in thermal equilibrium at all times during the evolution. 

\subsection{The role of the evolution model}

\begin{figure*}[!htb]
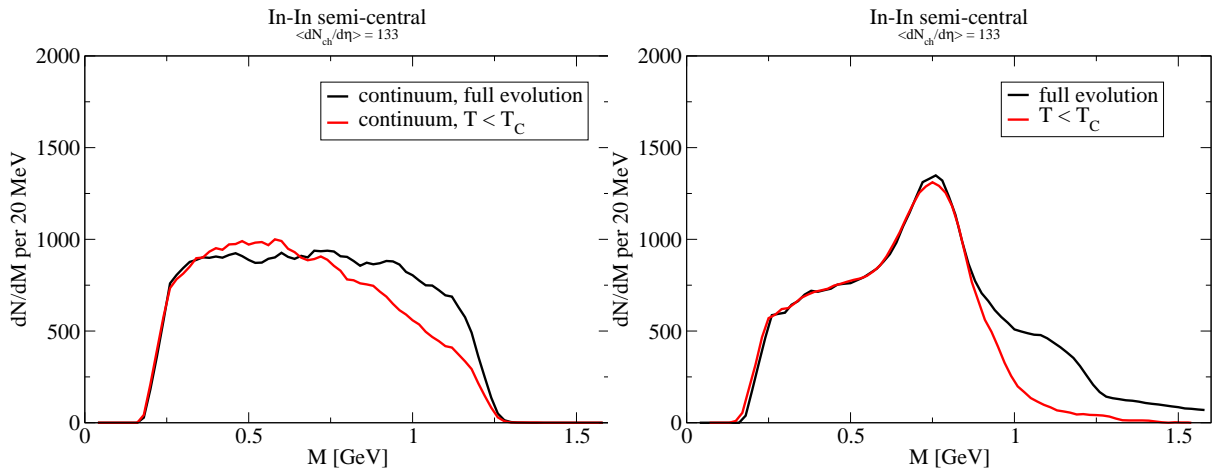

\epsfig{file=dileptons-InIn-RR2-continuum-hot-cold.eps, width=8cm}\epsfig{file=dileptons-InIn-RR2-fullspec-hot-cold.eps, width=8cm}
\caption{\label{F-EV}Comparison of the NA60 dimuon data \cite{NA60data} for integrated $p_T$ with different assumptions about spectral function and model evolution. Left panel: assuming $q\overline q$ spectral function at integrated $p_T$ for the full evolution ('hot') and the evolution below $T_C$ ('cold'). Right panel: Comparing the evolution below $T_c=170 {\rm MeV}$ with the evolution for all temperatures.
}
\end{figure*}

Looking at the $p_T$ integrated invariant mass spectrum, one is tempted to describe it as a vacuum $\rho$ peak on top of an almost symmetrically broadened structure, and naively one would infer that therefore the in-medium $\rho$ spectral function must show symmetrical broadening. However, it is rather easy to realize that this view has a major flaw: As Eq.~(\ref{E-Rate}) indicates, the emission rate is not given by the spectral function but by the spectral function multiplied by a thermal distribution. Thus, if the spectral function exhibits symmetric broadening, the corresponding rate cannot be symmetric for typical temperatures reached in heavy ion collisions but falls off sharply with increasing $M$. Furthermore, in the actual calculations, contributions from many volume elements with different flow and temperature are sampled and, indeed, the calculated distribution $d^2N/dMdp_T$ is anything but flat even for a symmetrically broadened spectral function.

It is a lucky but merely accidental coincidence that for a certain class of evolution scenarios a flat spectral function after folding with the fireball evolution and the acceptance returns a flat mass distribution over some range in $M$. Thus the data in this sense approximate the $(T,\rho_B)$) averaged lineshape of the spectral function. We illustrate this property in Fig.~\ref{F-EV}, left panel with the scenario called 'full evolution'. Here, we use the $q\overline{q}$ continuum spectral function (the strongest possible symmetrical broadening) throughout the whole evolution, i.e. above $T_C$ and below $T_C$. As aparent from the figure, an approximately flat signal is recovered after the acceptance folding in the mass region 0.25 GeV $< M <$ 1.2 GeV.  

However, this is not a general property of the acceptance but demands a specific type of evolution. To illustrate this point, we integrate radiation from the evolution model only from regions where $T<T_C$ (this could be pictured as the result of a long equilibration time $\tau_0$). In spite of the fact that the spectral function has the same continuum shape, the output is not flat in this case. It can clearly be observed that radiation from hot volume elements with $T>T_C$ is increasingly important for the contribution to invariant masses above the $\rho$ mass. 

This argument can be turned around: Even the maximal amount of broadening possible in a hadronic phase (i.e. the continuum spectral function) will not reproduce the data - if the normalization is such that the high $M$ region is described correctly, the yield overshoots the invariant mass region below the $\rho$ peak, if the low $M$ region is described correctly, there is not enough yield above the $\rho$ peak. 
We call this effect here therefore 'excess in the higher mass region' ($M \approx 0.9-1.5 {~\rm GeV}$).

We illustrate this point further in Fig.~\ref{F-EV}, right panel where the importance of radiation from regions with $T>T_C$ for the description of the invariant mass spectrum 
above the $\rho$ peak is clearly visible when the full hadronic spectral function is used.

This indicates that the most likely conclusion is that the data require radiation from a significant amount of matter with $T>T_C$ in order to account for the 
excess in the higher mass region. Our calculation of the medium evolution 
suggest therefore that those contributions are predominantly due to the early times
of the medium evolution.

The authors of \cite{vanHees:2006ng} argue differently and claim that  a further medium effect, namely contributions from four pion annihilation processes, accounts for the observed high mass excess. 
Since contributions to the spectrum from four pion annihilation would be dominated by the part of the evolution where the pion-chemical potential is large (and the rate is augmented by the pion fugacity factor $\exp{(4\mu_\pi/T})$), they have to assume that the process is important at lower $T \approx 120 {\rm MeV}$.
They furthermore need to have considerable chiral mixing of the free $V$ and $A$ correlators even at comparatively low temperatures.Their calculation estimates an upper limit on the in-medium four pion annihilation contribution to dilepton emission \cite{vanHees:2006ng,EPSILON}. 
This {\it upper} limit of the four pion annihilation with chiral mixing contribution to the spectrum is barely enough to account for the high mass $M>1 GeV$ region of the $p_T$ integrated data  in a medium evolution model with an initial temperature of $197 {~\rm MeV}$.

One might doubt on theoretical grounds that all of the assumption in
favor of this upper limit of the four pion annihilation contribution are realized.
But the question if the high mass excess is mainly due to emission at  high temperatures (accompanied by insignificant flow)  or low temperatures (accompanied by large flow) can be decided unambiguously experimentally. 

The important observation is that the $p_T$- dependence of the spectra at given $M$ with flow or temperature is different. Flow simply pushes spectra to higher $p_T$ and
reduces strength at lower $p_T$ accordingly whereas temperature dominated spectra are exponential without reducing strength at low $p_T$. 
$p_T$ spectra at large $M$ profit more from high flow than low mass spectra, for
temperature dependence the opposite is true.

Given only a comparison with the $p_T$ integrated data one can have similar spectra
from both mechanism since one can trade high $p_T$ contribution against
low $p_T$ contributions by suitable model adjustments. The two different mechanism can therefore lead to comparable predictions in the region $M>1{~\rm GeV}$ for all $p_T$ spectra.

This is different if one wants to describe low $p_T$ and high $p_T$ cuts at all $M$
consistently with one mechanism. For this not only an inclusion of the vacuum $\rho$ with the appropriate flow profile is necessary, but also the correct implementation
of the main contribution at high $M$. The four pion annihilation scenario has to predict
an access for the high $M$ spectrum that is different from what we find
find in our temperature dominated evolution.

Since low temperature (large flow) and high temperature (insignificant flow) alter the $p_T$ spectrum with a different
$M$ dependence, both mechanisms cannot lead to the same lineshape for 
different $p_T$ cuts, hence experimental discrimination is possible.

\section{Conclusions and outlook}

We have discussed a dynamical evolution model for the description of $158 {~\rm AGeV}$ In-In collisions and used it for the calculation of the dimuon spectrum measured
by NA60. 

We have shown that the model is in very good agreement with the  integrated $p_T$ NA60 spectrum \cite{NA60data}, if a spectral function including broadening effects and momentum dependence based on 
a $\Phi$-functional approach is applied. This not only provides a test for the evolution model, but also indicates that the vector spectral function 
found in nature must have similar features, i. e. similar $p_T$ dependence, no strong
mass shift and a moderate broadening of the $\rho$ meson. 

We do not claim to have identified the full physics of in-medium modifications of the $\rho$-meson with this
comparison since a scheme that incorporates chiral symmetry, self-consistent non-perturbative broadening effects, the full three-momentum dependence, finite temperature and density as well as the influence of heavier hadrons has not yet been achieved.

We furthermore provided calculations not only of the  $p_T$ integrated spectrum, but also
of high-$p_T$ and low-$p_T$ cuts. We expect these data to be published by the NA60 collaboration   soon. 

The low-$p_T$ analysis allows also a comparison with model calculation that do not
include a full momentum dependence of the spectral function. We calculated this spectrum therefore not only for the $\Phi$-derivable approach but also 
a thermal nuclear matter calculation \cite{Renk:2003hu} and 
for a $\rho$-meson spectral function in a chiral model calculation \cite{Klingl:1997kf}.The last two models also include further broadening via scattering off by baryons such an effect
should be reflected in an increase of the low $M<770 {~\rm MeV}$ contribution.
Substantial broadening is also predicted in QCD sum rule approaches with small or no
downward mass shifts \cite{BR}. 
The chiral model also includes contributions from a broadened in-medium $\omega$ and $\phi$. Considerable broadening of the $\phi$-meson should therefore
not only be reflected in integrated $p_T$, but should also be significant at low-$p_T$.

Discussing the importance of the evolution model we found that the $p_T$ integrated 
(even more the high-$p_T$) spectrum has significant contributions not only from in-medium vector $\rho$ contributions but also from the vacuum $\rho$. This is especially important in the  intermediate mass region $M \approx 0.6 - 0.9~{\rm GeV}$ where the vacuum rho has most of its spectral strength.

We also found that a purely hadronic evolution is unable to reproduce the data above
$1 {\rm GeV}$ invariant mass. This was a priori unexpected and is an interesting
hint that even in the rather small In-In system some amount of partonic evolution may contribute to the observed radiation. Clearly, the contribution is consistent with the radiation expected from a QGP, but without further evidence it would be premature to claim that a QGP has been observed in In-In collisions. Data which could help to pin down the hadronic freeze-out state (and hence the evolution) would clearly help to establish this conclusion more firmly.

It was show that this explanation can be discriminated experimentally from a scenario suggested in \cite{vanHees:2006ng} which tries to explain the high mass excess by four pion annihilations. The authors there have shown that they can describe the higher mass $M>0.9{~\rm MeV}$ part of the integrated $p_T$ 
data  in their evolution model if they use an upper bound for the contribution of four pion annihilations. It was also shown that considerable mixing has to be present
even at comparatively low temperatures in order to account for the higher mass excess in this scenario. 

We argue that this explanation can be discriminated from our explanation were
those higher mass contributions to the spectra are mainly due to the early and
hot evolution stages by considering lineshape in $M$ at different $p_T$ cuts. 
We provided calculations for the high and low $p_T$-cuts in our model. If the 
apparent good agreement with the shape of the preliminary data shown at Quark Matter 2005 \cite{SanjaQMTalk} is confirmed, a four pion annihilation scenario as the dominant source of the high mass excess can be ruled out experimentally. 

We also want to draw the reader's attention to an earlier study of us \cite{Renk:2006dt} where we argue that an additional experiment measurement of low mass di-muon production in Pb-Pb collisions at 158 AGeV with the same resolution as provided by NA60 for the In-In system can put further constrains on the theoretical description of the spectra.

\begin{acknowledgments}
We are grateful to Berndt M\"uller for many valuable comments, discussions and his support of this work.
We are grateful to Sanja Damjanovic for taking care of doing the NA60 acceptance simulations with our model input and (together with Hans Specht) for many interesting discussions. We are grateful to Carlos Lourenco for comments and discussions.
We would also like to thank Wolfram Weise and Norbert Kaiser for valuable comments and discussions. This work was supported by DOE grant DE-FG02-96ER40945
and the Alexander von Humboldt Foundation's Feodor Lynen Fellow program.
\end{acknowledgments}

\bibliography{u4}

\end{document}